\documentclass[onecolumn,superscriptaddress,nobibnotes,aps,prd,showpacs,nofootinbib,preprintnumbers]{revtex4}

\RequirePackage[english]{babel}
\RequirePackage[latin1]{inputenc}
\RequirePackage[T1]{fontenc}
\RequirePackage{mathrsfs}
\RequirePackage{amsmath}
\RequirePackage{amssymb}
\RequirePackage{amsbsy}
\RequirePackage{bm}
\usepackage[lofdepth,lotdepth]{subfig}
\usepackage{graphicx}
\usepackage{multirow} 
\usepackage{dsfont} 
\usepackage{color}
\def\rf#1{(\ref{#1})}

\def\de#1/de#2{\frac{\partial {#1}}{\partial {#2}}}
\newcommand{\R}{\mathcal{R}}

\newcommand{\h}{\mathcal{H}}
\newcommand{\N}{\mathcal{N}}
\newcommand{\M}{\mathcal{M}}
\newcommand{\ba}{\begin{eqnarray}}
\newcommand{\ea}{\end{eqnarray}}
\newcommand{\be}{\begin{equation}}
\newcommand{\ee}{\end{equation}}

\newcommand{\lp}{\left(}
\newcommand{\rp}{\right)}

\begin{document}

\title{A dynamical system analysis of hybrid metric-Palatini cosmologies}

\author{Sante Carloni}
\email{sante.carloni@tecnico.ulisboa.pt}
\affiliation{Centro Multidisciplinar de Astrofisica - CENTRA,
Instituto Superior Tecnico - IST,
Universidade de Lisboa - UL,
Avenida Rovisco Pais 1, 1049-001, Portugal}

\author{Tomi Koivisto}
\email{tomi.koivisto@nordita.org}
\affiliation{Nordita, KTH Royal Institute of Technology and Stockholm 
University, Roslagstullsbacken 23, SE-10691 Stockholm, Sweden}

\author{Francisco S. N. Lobo}
\email{fslobo@fc.ul.pt}
\affiliation{Instituto de Astrof\'{\i}sica e Ci\^{e}ncias do Espa\c{c}o, Faculdade de
Ci\^encias da Universidade de Lisboa, Edif\'{\i}cio C8, Campo Grande,
P-1749-016 Lisbon, Portugal}

\begin{abstract}

The so called $f(X)$ hybrid metric-Palatini gravity presents a unique viable generalisation of the $f(R)$ theories within the metric-affine formalism. Here the cosmology of the $f(X)$ theories is studied using the dynamical system approach. The method consists of formulating the propagation equation in terms of suitable (expansion-normalised) variables as an autonomous system. The fixed points of the system then represent exact cosmological solutions described by power-law or de Sitter expansion. The formalism is applied to two classes of $f(X)$ models, revealing both standard cosmological fixed points and new accelerating solutions that can be attractors in the phase space. In addition, the fixed point with vanishing expansion rate are considered with special care in order to characterise the stability of Einstein static spaces and bouncing solutions.

\end{abstract}

\preprint{NORDITA-2015-89}
\pacs{}
\date{\today}
\maketitle

\section{Introduction}

The so called hybrid $f(X)$ class of modified theories of gravity consists of the superposition of the metric Einstein-Hilbert Lagrangian
$R$ with an $f({\cal R})$ term constructed \`{a} la Palatini \cite{Harko:2011nh}. This interpolation between the non-equivalent metric and Palatini approaches has revealed interesting possibilities. For instance, in the formulation of the dynamically equivalent scalar-tensor representation, it was shown that the theories contain a ``light'', long-range scalar field, which is able to modify the cosmological and galactic dynamics while leaving the Solar System unaffected \cite{Harko:2011nh,Capozziello:2013uya}.
This feature has led to investigations of the cosmology in $f(X)$ theories \cite{Capozziello:2012ny,Capozziello:2012qt,Capozziello:2013yha,Borowiec:2014wva,Lima:2014aza,Lima:2015nma}, but a systematic mapping of its phase space is still missing. It is this gap we aim to fill in the paper at hand.

From a theoretical point of view, the $f(X)$ class of theories enjoys a similar special status amongst the more general hybrid metric-Palatini theories as the $f(R)$ theories within the narrower framework of purely metric gravity. To wit, when one excludes theories which present ghost-like, superluminally propagating and otherwise pathological degrees of freedom, there is evidence\footnote{In particular, we can exclude the more general cases of $f(R,\R)$ studied in \cite{Tamanini:2013ltp} and various second, fourth and sixth order derivative gravity theories resulting from considering more general curvature invariants constructed from the metric and the two available connections \cite{Koivisto:2013kwa}.} that the $f(X)$ class of theories is singled out as the only viable form of an action one can construct using the metric, its derivatives (and thus the metric Levi-Civita connection) and an independent ''Palatini connection''  \cite{Koivisto:2013kwa}. The underlying reason is \cite{Amendola:2010bk} that in the special case of $f(X)$ actions the higher derivatives in the gravity sector, that otherwise would result in an Ostrogradskian instability, can be separated into a scalar mode. Indeed the $f(X)$ gravity, where $X=R+ T$, represents a generic case within the one-parameter family of the Algebraic Scalar-Tensor theories, at one end of which lies the pure Palatini $f(\R)$ (wherein the field is a function of the stress energy trace $T$) and at the other end the pure metric $f(R)$ (where the field is a function of the metric curvature $R$) \cite{Koivisto:2009jn}. Furthermore, the propagating degrees of freedom have proven to be healthy also on curved backgrounds \cite{Capozziello:2012ny,Lima:2014aza}, and concerning the Cauchy problem, it was shown that in this class of theories the initial value problem can always be well-formulated and well-posed depending on the adopted matter sources \cite{Capozziello:2013gza}. These considerations provide compelling motivation for further exploration of these particular theories.

An immediate application is to consider gravitational alternatives to dark energy. As highlighted above, the hybrid theories are promising in this respect as they can naturally avoid the local gravity constraints but modify the cosmological dynamics at large scales. This is simply because as a scalar-tensor theory, the hybrid $f(X)$ gravity is characterised by an evolving Brans-Dicke coupling\footnote{To be precise, we have $\omega_{BD}=3\phi/(2\phi-2\Omega_A)$, where $\Omega_A=1$ for our case. This normalisation is chosen to produce the observed value of the Newton's constant today. The pure Palatini case is $\Omega_A \rightarrow 0$, and the purely metric limit is recovered at $\Omega_A\rightarrow \infty$.}, which allows to introduce potentially large deviations from General relativity in the past (and future) as long as the coupling at the present epoch is strong enough to hide the field from the local gravity experiments. In contrast, in the metric $f(R)$ models the Brans-Dicke coupling is a finite constant and one needs to invoke some of the various ''screening mechanisms'' \cite{screening} (workings of which remain to be studied in the hybrid theories) in order to reconcile the Solar System experiments with cosmology. Such dark energy dynamics that drive the theory towards its general relativistic limits today have indeed been realised in several specific models \cite{Harko:2011nh,Capozziello:2012ny,Lima:2014aza,Lima:2015nma}. A preliminary phase space analysis confirmed the existence of such de Sitter attractor solutions, and also other analytic solutions were presented in Ref. \cite{Capozziello:2012ny} as well as in Ref. \cite{Borowiec:2014wva}, there using a N\"other symmetry technique. A designer approach was employed in Ref.\cite{Lima:2014aza} to reconstruct precisely the standard $\Lambda$CDM expansion history by a nontrivial $f(X)$ model, and finally, two families of models were constrained by confronting their predictions with a combination of cosmic microwave background, supernovae Ia
and baryonic accoustic oscillations background data \cite{Lima:2015nma}. At this point, it can be useful to have a global view of the possible cosmological solutions from the perspective of dynamical system analysis.

Cosmological perturbations have been also analysed in these models up to the linear order \cite{Harko:2011nh,Capozziello:2012ny,Lima:2014aza}, and the results imply that the formation of large-scale in the aforementioned dark energy cosmologies is viable though exhibits subtle features that might be detectable in future experiments. The full perturbations equations were derived and at their Newtonian limit, describing the observable scales of the matter power spectrum, the growth of matter overdensities was shown to be modified by a time-dependent effective fifth force that will modify the red-shift evolution of the growth rate of perturbations \cite{Harko:2011nh,Capozziello:2012ny}. In addition, numerical studies of the perturbations imply that the difference of the gravitational potentials can exhibit oscillations at higher redshifts even when the background expansion and the full lensing potential are indistinguishable from the standard $\Lambda$CDM predictions \cite{Lima:2014aza}. Such features could potentially be observed in cross-correlations of the matter and lensing power spectra, but detailed calculations of the cosmic microwave background anisotropies and other perturbation observables remain to be carried out.

At an effective level, the $f(X)$ modifications involve both (the trace of) the matter stress energy and (the Ricci scalar of) the curvature, and from this point of view it appears 
appealing to speculate on the possible relevance of these theories to both the problems of dark energy and dark matter, in a unified theoretical framework and without distinguishing {\it a priori} matter and geometric sources \cite{Capozziello:2013uya}. Various aspects of dark matter phenomenology from astronomical to galactic and extragalatic scales have been considered in the presence of $f(X)$ corrections to gravity with some promising results \cite{Capozziello:2012qt,Capozziello:2013yha,Borka:2015vqa}. The generalised virial theoreom can acquire, in addition to the contribution from the baryonic masses, effective contributions of geometrical origin to the total gravitational potential energy, which may account for the well-known virial theorem mass discrepancy in clusters of galaxies \cite{Capozziello:2012qt}. In the context of galactic rotation curves, the scalar-field modified relations between the various physical quantities such as tangential velocities of test particles around galaxies, Doppler frequency shifts and stellar dispersion velocities were derived \cite{Capozziello:2013yha}. More recently, observational data of stellar motion near the Galactic centre was compared with simulations of the hybrid gravity theory, which turned out particularly suitable to model star dynamics \cite{Borka:2015vqa}. Yet, to promote the $f(X)$ theory into a convincing alternative to particle dark matter, one should also produce the cosmological successes of the $\Lambda$CDM model without the CDM component.  

In this paper we study the cosmological solutions of the theory in terms of dynamical system analysis. Cosmological applications of the phase space techniques have a long history, dating back to at least the early studies of chaos in the framework of cosmology by Belinsky \cite{Belisnky} and Misner \cite{Misner}, and were developed in particular by Ellis and Wainwright (see \cite{Dynamical} and references therein). The nowadays standard framework of constructing the phase space is based on defining expansion-normalised variables representing the various (effective) contributions to the total energy density. Besides mathematical efficiency this approach allows an easy physical interpretation of the phase space features as its variables assume the structure of cosmological parameters. The method has been used with great success in the analysis of anisotropic cosmologies \cite{Dynamical}, and has been applied e.g. to models with nonstandard equation of state \cite{Ananda:2006gf}, and more recently exploited to analyse a number of modified gravity models\footnote{To highlight the power of the method, we could mention that subjected to its scrutiny, the rather involved string-inspired set-up in an anisotropic spacetime filled by a massive vector field and a scalar field, and featuring their both conformal and disformal couplings, yielded to complete and detailed uncovering of its phase space \cite{fede}.} \cite{SanteDynSys}. In the context of $f(X)$ gravity, the stability of the Einstein universe, a particular static fixed point in the phase space of the theory, was analysed in great detail using the equivalent scalar-tensor description \cite{Boehmer:2013oxa}, and matter-dominated and de Sitter fixed points have been checked also \cite{Capozziello:2012ny}. Here we will set-up the phase space directly in terms of the curvature quantities, with the immediate advantage
of a more direct interpretation of the relation between the cosmic phenomenology and the form of the fundamental function $f(\R)$, and moreover allowing to take full advantange of the conformal relation between the metric and Palatini curvatures and resulting in a more efficient and generally applicable formulation of the dynamical system.    

This paper is outlined in the following manner: In Section \ref{SecII}, we present the basic formalism for the hybrid $f(X)$ gravity, namely, the action and the field equations and adapt them to cosmology. In Section \ref{SecIII}, we set up the dynamical system, which essentially consists of defining suitable variables and deriving their propagation equations. We then analyse the phase space structure in several families of particular models. In Section \ref{SecIV}, we reformulate the system in order to analyse the fixed points with $H=0$, that reside in the asymptotic regime of the phase space considered in the previous section. In Section \ref{conclusion}, we discuss our results and conclude.

As the reader may have noticed, we have set $8\pi G=1$.
Natural units ($\hbar=c=k_{B}=8\pi G=1$) will be used throughout this paper, and Greek indices run from 0 to 3. We use the $(-,+,+,+)$ signature and the Riemann tensor is defined by
\begin{equation}
{R}^{\beta}_{\;\;\alpha\mu\nu}
=\partial_{\mu}\Gamma_{\nu\alpha}{}^{\beta} - \partial_{\nu}\Gamma_{\mu\alpha}{}^{\beta} +
\Gamma_{\mu\delta}{}^{\beta}\Gamma_{\nu\alpha}{}^{\delta}-\Gamma_{\nu\delta}{}^{\beta}\Gamma_{\mu\alpha}{}^{\delta}\;,
\end{equation}
where the $\Gamma_{\mu\nu}{}^{\alpha}$ are the coefficients of a linear dynamical connection. 
The Ricci tensor is obtained by contracting the {\em first} and the {\em third} indices via the metric $g_{\mu\nu}$, i.e., ${R}_{\mu\nu}={R}^{\alpha}{}_{\mu\alpha\nu}$.

\section{Basic equations}\label{SecII}
The four dimensional action\footnote{In this setting $f$ contains a  coupling constant, but since it will be integrated in one of the variables, its presence will be irrelevant for the following discussion.} of hybrid metric-Palatini gravity is given by \cite{Harko:2011nh}
\begin{equation} \label{eq:S_hybrid}
S= \frac{1}{2}\int {\rm d}^4 x \sqrt{-g} \left[R + f(\R) + 2L_m(g^{\mu\nu}, \psi)\right]\,,
\end{equation}
where $L_m=L_m(g^{\mu\nu}, \psi)$ is the matter Lagrangian, coupling thes matter fields $\psi$ minimally to the metric $g_{\mu\nu}$,
$R$ is the Einstein-Hilbert term, $\R  \equiv g^{\mu\nu}\R_{\mu\nu} $ is
the ''Palatini curvature'', given by the contraction of the tensor $\R_{\mu\nu}$ that is generated by 
an {\it a priori} independent connection $\hat{\Gamma}^\alpha_{\mu\nu}$  as
\begin{equation}
\R_{\mu\nu} \equiv \hat{\Gamma}^\alpha_{\mu\nu ,\alpha} -
\hat{\Gamma}^\alpha_{\mu\alpha , \nu} +
\hat{\Gamma}^\alpha_{\alpha\lambda}\hat{\Gamma}^\lambda_{\mu\nu}
-\hat{\Gamma}^\alpha_{\mu\lambda}\hat{\Gamma}^\lambda_{\alpha\nu}\,.
\end{equation}
Varying the action (\ref{eq:S_hybrid}) with respect to the metric, one obtains the following gravitational field equation  
\be
\label{efe}
 G_{\mu\nu} +F(\R)\R_{\mu\nu}-\frac{1}{2}f(\R)g_{\mu\nu} = T_{\mu\nu}\,,
\ee
with the notation $F(\R) \equiv f'(\R)$ and the usual definition of the matter stress-energy tensor 
\begin{equation}
T_{\mu \nu}=-\frac{2}{\sqrt{-g}}\frac{\delta(\sqrt{-g}\,L_m)}{\delta(g^{\mu\nu})} \,.
\end{equation}    
The field equations for the connection imply that the independent connection is compatible with the conformal metric
$\hat{g}_{\mu\nu}=F(\R)g_{\mu\nu}$ and the relation
\ba
\label{ricci} \R_{\mu\nu} & = & R_{\mu\nu} +
\frac{3}{2}\frac{1}{F^2(\R)}F(\R)_{,\mu}F(\R)_{,\nu}
  - \frac{1}{F(\R)}\nabla_\mu F(\R)_{,\nu} -
\frac{1}{2}\frac{1}{F(\R)}g_{\mu\nu}\Box F(\R)\,. 
\ea 
connects the Palatini and the metric Ricci tensors.

Here we will focus on the cosmology of the hybrid models. We consider the homogenous and isotropic cosmologies  represented by the Friedmann-Lema\v{\i}tre-Robertson-Walker (FLRW) metric 
\begin{equation}
{\rm d}s^2=-{\rm d}t^2+a(t)^2\left[\frac{{\rm d}r^2}{1-kr^2}+r^2\left({\rm d} \theta^2+\sin^2\theta {\rm d}\varphi^2\right)\right]\,.
\end{equation}
The Hubble rate $H$ of the scale factor $a$ is defined as $H=\dot{a}/a$, where an overdot denotes a derivative with respect to the time coordinate $t$.
The Friedmann equations for this Hubble rate, derived from the above general equations, can be now written as \cite{Capozziello:2012ny}
\begin{align}
&\rho_{\rm eff}=3H^2 -3 k \label{cosmEqOrig1} \,, \\
&p_{\rm eff}=-2\dot{H}-3H^2+3k\label{cosmEqOrig2}\,,
\end{align}
where $\rho_{\rm eff}$ and $p_{\rm eff}$ are effective energy density and pressure given by
\begin{eqnarray}
\rho_{\rm
eff}&=&\mu - \Bigg\{\frac{1}{2}f(X)-\frac{3}{2}
\left[F''(X)-\frac{(F'(X))^2}{F(X)}\right]\dot{X}^2
     -\frac{3}{2}F'(X)\left(\ddot{X}+\dot{X}H\right)
 -3F(X)\left(\dot{H}+H^2 \right)\Big\}
\,,   \label{rhoeff} \\
p_{\rm eff}&=&p + \Bigg\{\frac{1}{2}f(X)-\frac{1}{2}F''(X)\dot{X}^2
     -\frac{1}{2}F'(X)\left(\ddot{X}+5\dot{X}H\right)
 -F(X)\left(\dot{H}+3H^2 \right)\Big\}\,,
 \label{peff}
\end{eqnarray}
respectively, where $\mu$ and $p$ are the energy density and pressure of a perfect fluid, and $X$ is defined as
\begin{equation} \label{TraceFE}
X \equiv T +  R=  F(\R)\R -2f(\R)  \,,
 \end{equation}
where we have used the trace of Eq.\rf{efe} in the last equality. Note that in the pure Einstein gravity one has $X=0$, and the curly bracket terms simply vanish:
this formulation of the cosmological equations has the advantage of making explicit the deviation of this class of theories from General Relativity and placing in evidence the scalar degree of freedom introduced by the Palatini term of the action \rf{eq:S_hybrid}.

As the next step, we take advantage of the conformal relation between the two geometries (\ref{ricci}) that naturally appear in the theory.
Indeed, the equations (\ref{cosmEqOrig1}) and (\ref{cosmEqOrig2}) can be simplified with a suitable choice of variables that exploits in each occasion the optimal quantity associated with either $\hat{g}_{\mu\nu}$ or $g_{\mu\nu}$. Defining the auxiliary function $A=\sqrt{F(\R)} \,a(t)$, one can write
\begin{eqnarray}
\h &=& \frac{\dot{a}}{a}+\frac{\dot{F}}{2 F} =\frac{\dot{A}}{A}, \label{A-R}\\
 \R &=& 6F\left[\frac{{A}^{\dagger\dagger}}{A} +\left(\frac{A^\dagger}{A}\right)^2+\frac{k}{A^2} \right], \label{A-R2}
\end{eqnarray}
where the symbol $\dagger$ indicates the derivative with respect to $\tau$, defined by $d\tau=\sqrt{F(\R)} dt$. Using these quantities defined in Eqs.(\ref{A-R})-(\ref{A-R2}) to rewrite the field equations \rf{efe}, they assume the following form:
\begin{align}
& \left(\frac{\dot{a}}{a}\right)^2 + \frac{k}{a^2}(1+F)+ F \h^2-\frac{1}{6}\left(F\R-f\right)-\frac{\mu}{3}=0\,, \label{fr1} \\
&\frac{\ddot{a}}{a}-\left(\h^2 +\frac{k}{a^2}\right) F + \frac{f}{6}+\frac{1}{6}\lp\mu+3p\rp =0 \,, \label{fr2} \\
& \dot{\mu}+3\frac{\dot{a}}{a}(1+w)\mu=0\,,\label{fr3}
\end{align}
where we have used the equation of state $p=w\mu$, and for notational simplicity, we have dropped the dependence of $f$ and $F$ of $\R$. 
This formulation of Eqs. (\ref{cosmEqOrig1}) and (\ref{cosmEqOrig2}) can be  used to deduce some basic features of the hybrid metric-Palatini theories. For example, Eq. \rf{fr2} shows clearly that the interaction between the Palatini and metric part of the theory can generate an increase of the expansion rate for a flat and spatially closed universe ($F$ must be definite positive for the transformation $\hat{g}_{\mu\nu}=F(\R)g_{\mu\nu}$ to make sense), whereas the nonminimal coupling of the Palatini curvature can have the same effect only if the function $f$ is negative. This means that a sufficient condition for the model to generate cosmic acceleration is $f<0$ and $k=0,1$ (at least when the geometrical contribution dominates the matter sources). Similar considerations can be used in other frameworks such as in the context of (a homogenous and isotropic) gravitational collapse. 

In the following we will use the above equations to implement the dynamical system analysis. Differently from previous similar attempts, we will not rely on the definition of an auxiliary scalar field to perform this task, but we will only use Eqs. (\ref{fr1})-(\ref{fr3}).

\section{General dynamical systems formalism}\label{SecIII}

In this section, we will construct the dynamical system for the hybrid models, which consists of essentially two steps: (i)  the definition of suitable variables, and (ii) the derivation of the propagation equations for these variables. Basically this is rewriting of the system of differential equations as a first order system of the chosen different variables. The fixed points will thus correspond to particular exact solutions of our original Friedmann equations (\ref{cosmEqOrig1}) and (\ref{cosmEqOrig2}).

\subsection{General formalism}
Let us define the variables
\begin{align}\label{DynVar}
&X=\frac{\h}{H}\,,\qquad Y=\frac{\R}{6H^2}\,,\qquad Z=\frac{f}{6H^2}\,, \qquad\Omega= \frac{\mu}{3H^2}\,,\qquad K=\frac{k}{a^2H^2}\,.
\end{align}
The interpretation of these variables is straightforward, as each represents the corresponding quantity in units of the expansion rate.
We also introduce a dimensionless time variable: the logarithmic time $\N=\ln a$. Note that with this time variable we are implicitly imposing that the orbit of the phase space will represent only monotonically expanding or contracting cosmologies. We will focus on the possibility of solutions crossing $H=0$ in the following section.

In terms of the variables (\ref{DynVar}), the cosmological equations can now be written as the following dynamical system:
\begin{align}\label{DynSysGen}
\begin{split}
& X_{,\N}=\frac{1}{2} X \left[2 Z-X-2\mathcal{F}\left(K+X^2\right)+(1+3 w) \Omega \right]-K+Y\,,\\
& Y_{,\N}=Y \left[2(1+Z)-2\mathcal{F}\left(K+X^2\right)+(1+3 w) \Omega +2
   (X-1) \mathcal{Q}\right]\,,\\
& Z_{,\N}=Z \left[2(1+Z)-2\mathcal{F}\left(K+X^2\right)+(1+3 w) \Omega\right]+2  Y(X-1)
  \mathcal{F}\mathcal{Q}\,,\\
&\Omega_{,\N}=\Omega  \left[2 Z-2\mathcal{F}\left(K+X^2\right)+(1+3 w) ( \Omega-1)\right]\,, \\
&K_{,\N}=K \left[2 Z-2\mathcal{F}\left(K+X^2\right)+(1+3 w) \Omega  \right]\,.
\end{split}
\end{align}
with the constraint
 \begin{equation}
 K [1+\mathcal{F}]+X^2 \mathcal{F}-Y \mathcal{F}-\Omega +Z+1=0\,.
 \label{constrainteq}
 \end{equation}
In the above equations, the quantities $\mathcal{F}= F(\R)$ and $\mathcal{Q}(\R)=F/(\R F')$ are only functions of $\R$,  and in order to close the system they have to be expressed in terms of the variables specified in Eq (\ref{DynVar}). Given the form of $f$, this can be done by noticing that $Z/Y$ is function of $\R$ only, i.e.,
 \begin{equation}\label{KeyEq}
 \frac{Z}{Y}=\frac{f(\R)}{\R}.
 \end{equation}
 Inverting this relation for $Y\neq0$, one obtains $\R=\R(Z/Y)$ and  the dynamical system can be closed. Therefore, as far as this inversion is possible, the phase space spanned by our dynamical variables (\ref{DynVar}) is fully described by the closed system of propagation equations \rf{DynSysGen}. The price to pay for this is the same that one finds in the application of a similar method to the metric $f(R)$ gravity \cite{Amendola:2006we,Carloni:2007br}, namely that the phase space can have singular surfaces (such as $Y=0$), which may have to be treated with great care. 
 
In the following we will use the constraint equation (\ref{constrainteq}) to eliminate the variable $\Omega$.  The independent equations of the system are
 \begin{align}
 \begin{split}\label{DynSysRed}
& X_{,\N}=\frac{1}{2} X \left\{\mathcal{F} \left[(3 w-1) \left(K+X^2\right)-(3 w+1) Y\right]+(1+3 w)(K+1)+3 (w+1) Z-2X\right\}-K+Y\,,\\
& Y_{,\N}=Y\left\{\mathcal{F} \left[(3 w-1) \left(K+X^2\right)-(3 w+1) Y\right]+(1+3 w)K+3 (w+1) (Z+1)+2 \mathcal{Q} (X-1)+2\right\}\,,\\ 
&Z_{,\N}=Z \left\{\mathcal{F} \left[(3 w-1) \left(K+X^2\right)-(3 w+1) Y\right]+(1+3  w)K+3(w+1) (Z+1)+2\right\}+2 \mathcal{F}\mathcal{Q}Y(X-1) \,,\\
&K_{,\N}=K \left\{\mathcal{F} \left[(3 w-1) \left(K+X^2\right)-(3 w+1) Y\right]+(1+3  w)(K+1)+3 (w+1) Z\right\}\,.   
\end{split}
\end{align}
Any other choice of independent dynamical variables could be made in principle, but  the one above is the only one that does not introduce divergences in the dynamical equations.
 
 The system above can then be analyzed with the standard tools of dynamical system analysis and in particular with the Hartmann-Grobmann theorem. Once the fixed points have been found, together with their stability, the solutions associated to the fixed points correspond to
 \begin{eqnarray}
&& a=a_0(t-t_0)^{1/\gamma}\,, \\
&& \mu=\mu_0(t-t_0)^{-3(1+w)/\gamma}\,,
\end{eqnarray}
where the constant $\gamma$ is obtained by expressing the modified Raychaudhuri equation \rf{fr2} in terms of the dynamical variables (the asterisk denoting their respective value at the fixed point)
\begin{equation}
\gamma=K_* \mathcal{F}_*+X_*^2 \mathcal{F}_* -\frac{1}{2}(1+3w) \Omega_*-Z_*-1\,. \label{Alpha}
\end{equation}
If $\gamma$ is equal to zero the formulae above do not hold and we obtain an exponential solution, which can be fully specified as well by substitution into the cosmological equations.

\subsection{Specific case I: $f =\chi\R^n$}
As an example let us look at the specific case $f=\chi\R^n$. We have, for $Y\neq0$,
\begin{equation}\label{closureRn}
 \frac{Z}{Y}=\chi\R^{n-1}\,,
\end{equation}
so that
\begin{equation}
 \mathcal{F}= \frac{nZ}{Y},\qquad  \mathcal{Q}= \frac{1}{n-1}\,,
 \end{equation}
It is important to point out here that in the case $\chi=0$ the \rf{closureRn} cannot be inverted. This result implies that we will always consider $\chi\neq0$ in our analysis\footnote{This conclusion is in general true for every type of $f(\R)$: if the coupling constant in $f$ is zero our entire construction cannot be applied.}.  The system \rf{DynSysRed} reduces to 
\begin{align} \label{SysRn}
\begin{split}
& X_{,\N}=\frac{1}{2} X \left\{\frac{nZ}{Y}\left[(3 w-1) \left(K+X^2\right)-(3 w+1) Y\right]+(1+3 w)(K+1)+3 (w+1) Z-2X\right\}-K+Y,\\
& Y_{,\N}=Y\left[(1+3 w)K+3 (w+1) (Z+1)+2 \frac{X-1}{n-1}+2\right]+Z \left[(3 w-1) \left(K+X^2\right)-(3 w+1) Y\right]\,,\\ 
&Z_{,\N}=Z \left\{\frac{nZ}{Y} \left[(3 w-1) \left(K+X^2\right)-(3 w+1) Y\right]+(1+3  w)K+3(w+1) (Z+1)+2 +\frac{2n}{n-1}(X-1)\right\},\\
&K_{,\N}=K \left\{\frac{nZ}{Y} \left[(3 w-1) \left(K+X^2\right)-(3 w+1) Y\right]+(1+3  w)(K+1)+3 (w+1) Z\right\}\,.
\end{split}
\end{align}
It is important to stress at this point that the system above is equivalent to \rf{DynSysRed} only if $Y\neq0$, so fixed points with $Y=0$ are excluded. This fact remains true even if no denominator containing $Y$ is present. The reason is that the \rf{KeyEq} only holds for  $Y\neq0$ and therefore  for $Y=0$ one cannot close the system. Note also that the \rf{SysRn} now includes two invariant submanifolds,  $Z=0$ and $K=0$. This result implies that only points with $Z=0$, $K=0$ (and $\Omega=0$) can be global attractors for the phase space.

The fixed points are determined setting to zero the $\N$-derivatives of the dynamical system variables. The finite phase space contains three fixed points. Their associated  solutions and stability can be found in Table \ref{T-Rn-PF}. We can distinguish two classes of fixed points depending on the value of the variable $Z$. Points with $Z=0$  ($\mathcal{B}, \mathcal{C}$) essentially represent states in which $f(\R)\approx0$ and therefore in which this model approximates General Relativity. 

It is also worth noting that the points $\mathcal{B}$ and $\mathcal{C}$, can be characterized by a negative value of $X$ i.e. has the peculiar characteristic to have negative "Palatini Hubble rate" while the physical universe is expanding or vice versa. 

All the solutions associated to the fixed points are the typical GR  Friedmann solutions already found in  \cite{Capozziello:2012ny,Borowiec:2014wva}. In addition, the coordinates of the associated fixed points reveal that GR-like solutions can only be realized $Z=0$, as expected.  Point $\mathcal{A}$ is associated to an exponential expansion whose time parameter $\lambda$  can be obtained from the trace of the gravitational field equations (\ref{TraceFE}). In the case of $f=\R^n$  this equation reduces to
\begin{equation}
(n-2)\R = T +  R\,.
\end{equation}
Then, in the case of an exponential solution in vacuum, $R$ is constant  and 
\begin{equation}
\R =  \frac{R}{n-2}.
\end{equation}
This immediately implies that $H=\h$ and substituting into the Friedmann equation one obtains $\lambda=0$ and
\begin{equation}\label{lambSol}
\lambda=\frac{1}{2\sqrt{3}}(n-2)^{\frac{n}{2 (n-1)}} [(4-n) n-2]^{\frac{1}{2-2 n}} \chi ^{\frac{1}{2-2 n}}\,.
\end{equation}
For a given $n$ this expression will give multiple solutions. This is a common occurrence in higher order gravity and it is also present in the hybrid metric-Palatini case. However, the problem arises of how to associate these values of $\lambda$ to the solutions in the fixed points. The case  $\lambda<0$  can be excluded if we consider only expanding cosmologies, and the case $\lambda=0$ because it implies $H=0$ that cannot be considered finite phase space. Further selection criteria can be obtained by checking the values of the dynamical variables at the fixed points. In the case of $\mathcal{A}$, one has $Y>0$ which implies $\R>0$. This result allows to exclude all the values of $\lambda$ associated to $\R<0$. Therefore one is left with the (unique) real positive value of Eq. \rf{lambSol} which requires $\R>0$. Unfortunately, no general expression for this value of $\lambda$ can be given explicitly, but its calculation is straightforward for a specific $n$.

Looking at the stability of the fixed points, there are only two possible attractors: $\mathcal{A}$ and $\mathcal{B}$. The last point behave as a sink only for specific values of $n$. Since, however,  these points do not have $Z=0$, $K=0$ they cannot be  global attractors. In fact, in general no fixed point can be a global attractor for the cosmology of this model. The  points with $Z=0$ are generally unstable for every value of the parameter $n$.  The exception is given by the point $\mathcal{B}$ for $1<n<2$. This result implies that GR-like Friedmann cosmologies are in general unstable in this model and only a special set of initial conditions  might lead to the (non global) attractor $\mathcal{B}$. Such feature can be interesting to model the onset of a dark era in this context.
\begin{table}[tbp] \centering
\caption{The fixed points and the solutions of the hybrid model with $f=\chi\R^n$. Here the value of lambda is given by the \rf{lambSol}. In addition, A stays for attractor, F$_{A}$ for attractive focus, R for repeller and S for saddle.  }
\begin{tabular}{cclllccc}
& & \\
\hline   Point &$(X, Y, Z, \Omega,K)$ &  Scale Factor  & Energy Density& Stability\\ \hline
& & \\
$\mathcal{A}$& $\left\{1,2,\frac{2}{n-2},0,0\right\}$& $a=a_0 \exp\left[\lambda(t-t_0)\right]$ & $\mu=0$&F$_A$\\
& & \\
\multirow{3}{*}{$\mathcal{B}$} &\multirow{3}{*}{$\left\{2-n,(n-1)(n-3),0,0, -1\right\}$}&  \multirow{3}{*}{$a=a_0 \left(t-t_0\right)$} & \multirow{3}{*}{$\mu=\mu_0 \left(t-t_0\right){}^{-3 (w+1)}$} & F$_A$~~ $1<n<2$\\
&&&&~~~~~~ $0\leq w\leq 1$ \\
 &&&&S otherwise\\
& & \\
&$\left\{\frac{1}{2} [-3 n(w+1)+3 w+5],\right.$&  \\ 
$\mathcal{C}$ & $\frac{1}{4} [3 n (w+1)-4] [3 n (w+1)-3 w-5],$& $a=a_0 \left(t-t_0\right)^\frac{2}{3(1+w)}$&$\mu=\mu_0(t-t_0)^{-2}$& S\\
&$\left.0,1,0\right\}$ \\
& & \\
 \hline
\end{tabular}\label{T-Rn-PF}\\
\end{table}

\subsection{Specific case II: $f=\alpha \frac{\R}{1+\beta\R}$}\label{HS}
As a further example, let us consider the case of the function $f=\alpha \R/(1+\beta\R)$, with $\alpha$ and $\beta$ generic constants. This model is inspired by the proposal of Hu and Sawicki \cite{Hu:2007nk} in the context of the metric $f(R)$ gravity and we will refer to it as ``hybrid HS'' model. As in that case the function $f$ is such that in expanding cosmologies for very high and very small values of the Hilbert-Einstein term $R$, $T$ is also very large or very small  and the Palatini curvature $\R$ is constant.  In particular with our choice of the values of the parameter, $\R$ becomes zero ($\alpha$ positive) or $-\frac{1}{2\beta}$ ($\alpha$ negative) for $R\rightarrow0$ and becomes $-\frac{1}{\beta}$ if $R\rightarrow\infty$. Since we also require that $F>0$ the constant $\alpha$ must be chosen positive. Thus effectively the function $f$ represents an effective cosmological constant term which appears in the late universe, but does not affect the early one. Note that the sign of this effective cosmological constant depends on the sign of $\beta$, so that if $\beta<0$ the effective cosmological constant in Eq. \rf{efe} is positive otherwise, if $\beta>0$, it is negative. We will choose, in the following, the first option.

It is also worth noticing that for this type of function $f$, in the limit $\R \rightarrow (0, \infty)$, $F$  becomes constant and $\h=H$ so that the equations effectively reduce to GR plus a cosmological constant. This is a characteristic property of the Hu-Sawicki  model in the context of metric $f(R)$-gravity and it is present also in the hybrid metric Palatini case.  

For the hybrid HS model we have, for $Y\neq0$,
\begin{equation}
 \frac{Z}{Y}=\frac{\alpha}{1+\beta \R}
\end{equation}
so that
\begin{equation}\label{FQHS}
\mathcal{F}= \frac{Z^2}{\alpha Y^2},\qquad  \mathcal{Q}= \frac{\alpha Y }{2(Z-\alpha Y)}, 
\end{equation}
Note that in the case $\beta=0$, $f(\R)$ becomes linear. This case is degenerate in terms of the dynamical system variables ($Y=Z$) and has to be treated separately\footnote{In fact upon closer inspection, it appears clear that if $F=const.$ then $A\propto a$ and $\h=H$ and the dynamics is governed only by the variable $\Omega$. The phase space is one dimensional and presents only a single, unstable, fixed point corresponding to a $t^{2/3(1+w)}$ solution.}. Using \rf{FQHS}, the system \rf{DynSysRed} becomes
 \begin{align}
 \begin{split}\label{SysHS}
& X_{,\N}=\frac{1}{2} X \left\{\frac{Z^2}{\alpha Y^2}\left[(3 w-1) \left(K+X^2\right)-(3 w+1) Y\right]+(1+3 w)(K+1)+3 (w+1) Z-2X\right\}-K+Y\\
& Y_{,\N}=\frac{Z^2}{\alpha Y}\left[(3 w-1) \left(K+X^2\right)-(3 w+1) Y\right]+Y\left[(1+3 w)K+3 (w+1) (Z+1)+  \frac{\alpha Y(X-1) }{Z-\alpha Y}+2\right]\,,\\ 
&Z_{,\N}=Z \left\{\frac{Z^2}{\alpha Y^2} \left[(3 w-1) \left(K+X^2\right)-(3 w+1) Y\right]+(1+3  w)K+3(w+1) (Z+1)+\frac{Z(X-1) }{Z-\alpha Y} \right\},\\
&K_{,\N}=K \left\{\frac{Z^2}{\alpha Y^2}\left[(3 w-1) \left(K+X^2\right)-(3 w+1) Y\right]+(1+3  w)(K+1)+3 (w+1) Z\right\}\,.
\end{split}   
\end{align}
The above system only admits the invariant submanifolds $Z=0$, $K=0$ and presents a singular subspace for $Z=\alpha Y$. The analysis of the phase space will be performed in the $Z>\alpha Y$ and $Z<\alpha Y$ part of the space with the regular techniques. We do not address here the behaviour of the orbits close to  $Z=\alpha Y$.

The system \rf{SysHS} admits four fixed points as indicated in Table \ref{TableHS}. Of these points, two ($\mathcal{A}$,  $\mathcal{B}$)  correspond to de Sitter solutions and the other two  have $Z=0$ and represent GR-like solutions.   

Again, the time constant for the de Sitter solutions can be calculated using the trace equation, rewritten in Eq. \rf{TraceFE}. However, for this case the solution for the Palatini curvature is given by
\begin{equation}
\R=-\frac{\alpha +24 \beta  \lambda ^2\mp\sqrt{\alpha(\alpha -48 \beta  \lambda ^2)} }{4 \beta
   \left(\alpha  +6 ^2 \lambda ^2\right)} ,
\end{equation}
which is not unique. Therefore, differently from the previous case and the $f(R)$ case,  here the de Sitter solutions are set apart by the value of the Palatini curvature, other than $R$. As in the previous example the specific value of $\lambda$ associated to a fixed point can be deduced via considerations on the type of evolution represented by the phase space and the coordinates of the fixed points. In particular,  $\lambda=0$ and $\lambda<0$   as well as the values of $\lambda$ for which $\R$ diverges have to be discarded. This leaves only two values of $\lambda$, given by
\begin{equation}\label{LambdaValuesHS}
\lambda_1=\frac{1}{2}\sqrt{-\frac{ 1+\alpha+\sqrt{\alpha(\alpha +1)}}{3\beta }}\,, \qquad \lambda_2=\frac{1}{2}\sqrt{\frac{\alpha+1-\sqrt{\alpha(\alpha +1)}}{3\beta }} \,,
\end{equation}
respectively. Assuming $\beta<0$, we have that, for $0<\alpha<1/3$, $\lambda_1$ is associated to $\R>0$ and $\lambda_2$ to $\R<0$. Instead, for $\alpha>1/3$, $\lambda_1$ is associated to $\R<0$ and $\lambda_2$ to $\R>0$. As a consequence, the nature of the fixed points depends on the value of $\alpha$ other than their coordinates. Specifically, points $\mathcal A$ and $\mathcal B$ that have $Y>0$ are characterised by $\lambda_1$ for $0<\alpha<1/3$ and $\lambda_2$ for $\alpha>1/3$. 

The calculation for the stability for these fixed points is straightforward. The points $\mathcal{A}$ and $\mathcal{B}$ are always stable. The phase space present no other possible attractor. Since global attractors can only be flat GR solutions ($Z=0$, $K=0$), in spite of the generation of an effective cosmological constant discussed in the beginning, our analysis shows set of initial conditions for which a de Sitter expansion is not realised. In other words, the phenomenon of approaching to a future dominance of an effective cosmological term is not obvious as our initial reasoning seems to indicate. An escape from this conclusion could be the presence of some asymptotic ($H=0$) attractor. We will see in the following that this is not the case. 

Notice also that in this model there is no GR-Friedmann past attractor.  However considering also the stability of point $\mathcal{D}$ which is related to the classical Friedmann solution we can conclude that GR-like Friedmann cosmologies are always unstable.

\begin{table}[tbp] \centering
\caption{The fixed points and the solutions of the case with $f=\alpha \frac{\R}{1+\beta\R}$. The value of the parameter $\lambda$ is determined by Eq. \rf{LambdaValuesHS}. 
Here $A$ stands for an attractor, R for a repeller, $S$ for a saddle, F$_A$ for an attractive focus and F$_R$ for a repulsive focus. The stability of point $\mathcal{J}$ is given only for the case $w=0$.}
\begin{tabular}{ccccl}
& & \\
\hline   Point &$(X, Y, Z, \Omega,K)$ &  Scale Factor  & Energy Density & Stability \\ \hline
& & \\
 \multirow{2}{*}{$ \mathcal{A}$} & \multirow{2}{*}{ $\left[1,2,2 \alpha -2 \sqrt{\alpha  (\alpha +1)},0, 0\right]$}&  \multirow{2}{*}{ $a=a_0 \exp\left[\lambda(t-t_0)\right]$} &  \multirow{2}{*}{ $\mu=0$} &  F$_A$~~$ 0<\alpha < 49/32$\\ 
& & &&A~~otherwise\\
& & \\
$\mathcal{B}$&  $\left[1,2,2 \alpha +2 \sqrt{\alpha  (\alpha +1)},0, 0\right]$&  $a=a_0 \exp\left[\lambda(t-t_0)\right]$ & $\mu=0$ & F$_A$  \\ 
& & \\
$\mathcal{C}$& $\left[3,8, 0,0, -1\right]$&  $a=a_0 \left(t-t_0\right)$& $\mu=\mu_0(t-t_0)^{-3(1+w)}$ &S \\ 
 & &  \\ 
 $\mathcal{D}$& $\left[4+3w,\frac{1}{2} (3 w+4) (3 w+7),0,1,0\right]$&$a=a_0\left(t-t_0\right)^{\frac{2}{3(1+w)}}$&$\mu=\mu_0(t-t_0)^{-2}$& S\\
 & & \\
\hline
\end{tabular}\label{TableHS}
\end{table}

\section{Fixed points with $H=0$}\label{SecIV}
The dynamical system formulation above is not the only one that can be constructed for the cosmological equations (\ref{fr1})-(\ref{fr2}). Another possibility is to use the following variables 
\begin{align}\label{BarVar}
&\bar{X}=\frac{H}{\h},\qquad \bar{Y}=\frac{\R}{6\h^2},\qquad \bar{Z}=\frac{f}{6\h^2}\,,\qquad
\bar{\Omega}= \frac{\mu}{3\h^2},\qquad \bar{K}=\frac{k F}{A^2\h^2}\,,
\end{align}
where $\h=\sqrt{F(\R)}A^\dagger/A=\dot{A}/A$, together with the time variable $\mathcal{M}$ defined in such a way that, for a given dynamical system variable $P$, $P_{,\mathcal{M}}=\h^{-1}P_{,\tau}$.  The previous set of variables differs from this case, as here one is able to see fixed points characterized by $H=0$ which are in the asymptotic regime of the previous phase space.  Using the \rf{BarVar} we are able  to characterize the stability of Einstein static universes in the hybrid models and to  gain information on bounce  ($H=0$, $\dot{H}>0$)  phenomenology or other changes in the sign of $H$ typical of higher order theories of gravity \cite{Carloni:2005ii}. 

Before proceeding with the analysis, it is important to stress that the new time coordinate $\bar{\mathcal{M}}$ differs in sign from  $\mathcal{M}$ whenever $\bar{X}<0$. This fact implies that in treating the stability, fixed points for which $\bar{X}<0$ will have different stability (e.g. attractors could correspond to repellers for given values of the parameters) with respect to the ones obtained with the variables of the previous section. As we will see in the example we considered the stability of the fixed points does not change, but this might be the case in other modelts. It is clear however that, since $H$ is related to the physical time the ``correct'' stability is the one obtained in the previous section. 

\subsection{Dynamical system equations}

The dynamical system equations in this case read
\begin{align}
& \bar{X}_{,\M}=(\bar{K}+1) \mathcal{F}+\bar{X} (\bar{K}-\bar{Y}+1)-\frac{1}{2} (1+3 w) \Omega - \bar{Z}\,,\\
& \bar{Y}_{,\M}=2 \bar{Y} [\bar{K}+\bar{X} -\bar{Y}-(\bar{X}  -1)\mathcal{Q}+1]\,,\\ 
& \bar{Z}_{,\M}=2\bar{Z}[(\bar{K}+\bar{X} -\bar{Y}+1)  -(\bar{X} -1) \bar{Y} \mathcal{F} \mathcal{Q}\,,\\
&\bar{\Omega}_{,\M}=\Omega  [2 \bar{K}-(3 w+1) \bar{X} -2 \bar{Y}+2]\,, \\
&\bar{K}_{,\M}=2\bar{K}(1+\bar{K}- \bar{Y})\,,
\end{align}
with the constraint
 \begin{equation}
(1+\bar{K} -\bar{Y})\mathcal{F}+\bar{K}+\bar{X}^2-\bar{\Omega} +\bar{Z}=0\,,
 \end{equation}
 and the evolution equation for $ \h$
 \begin{equation}\label{eqD}
 \h_{,\M}=- (1+\bar{K}+\bar{X}-\bar{Y})\h\,,
 \end{equation}
 which is decoupled from the other dynamical equations.
 
 Analogously, in the following, we will eliminate the variable $\bar{\Omega}$ for consistency. In this way the independent equations of the system are the following
\begin{align}
& \bar{X}_{,\M}=X\left[K-Y+1-\frac{1}{2} (1+3 w) X\right] -\frac{1}{2} \{K (1+3 w)+3 (1+w) Z-[(1-3 w) (K+1)+(1+3 w)Y] \mathcal{F} \},\\
& \bar{Y}_{,\M}=2 Y [K+X-Y-(X -1)\mathcal{Q}+1],\\ 
& \bar{Z}_{,\M}=2[(K+X-Y+1) Z-(X-1) Y \mathcal{F} \mathcal{Q}],\\
&\bar{K}_{,\M}=2\bar{K}(1+\bar{K}- \bar{Y}).
\end{align} 
The solutions associated to the fixed points in this case are given by equation is Eq. \rf{eqD} rather than the Raychaudhuri equation (which is used only when $\h(t)$ has been determined in the fixed point). We have 
\begin{equation}
\dot{H}=-\left[\bar{X}_*^2+\bar{Z}_* -(1+\bar{K}_*)\mathcal{F}_*+\frac{1}{2}(1+3w)\bar{\Omega}_*\right]\h,
\end{equation}
where the asterisk means, as usual, that the variable is evaluated in the fixed point. From this result one can derive $a$ and $\mu$. 

In the following we will repeat the analysis of the $f=\chi\R^n$ and the hybrid HS model using these variables.
 
\subsection{The case $f=\chi\R^n$ in the barred variables}\label{RnBar}
In this case the general system reduces to 
\begin{align}\label{DynSysBarRn}
& \bar{X}_{,\M}=X\left[K-Y+1-\frac{1}{2} (1+3 w) X\right] -\frac{1}{2} [K (1+3 w)+3 (1+w) Z- \frac{nZ}{Y}[(1-3 w) (K+1)+(1+3 w)Y] ,\\
& \bar{Y}_{,\M}=2 Y [K+X-Y-\frac{X -1}{n-1}+1],\\ 
& \bar{Z}_{,\M}=2Z\left[K+X-Y+1 - \frac{n}{n-1}(X-1) \right],\\
&\bar{K}_{,\M}=2\bar{K}(1+\bar{K}- \bar{Y}).
\end{align}
The fixed points are given in Table \ref{T-Rn-PF_Bar} with their associated solutions and stability. Note that there is a one to one correspondence to the solutions found in the previous section. In place of these points, there is now one additional fixed point, $\bar{\mathcal{D}}$, which represents a static universe (since $X=0$, in these points $H=0$\footnote{To be completely precise, $H=0$ does not necessarily require $\bar{X}=0$. In fact, $\bar{X}$ can be finite at the bounce if $\h=0$ or $F=0$ - such solutions have indeed been found to exist in the Palatini-type ($\Omega_A\rightarrow 0$) theories \cite{gonzalo}. However in the first case the \rf{A-R} implies $F=const.$ which means that the Palatini term is linear. We have already seen that this case is degenerate for the dynamical system point of view. In the second case ($F=0$), since the metrics $g$ and $\hat{g}$ are not related anymore, the formulation of the theory used in our approach  looses meaning }). Since the point $\bar{\mathcal{D}}$ is characterized by $Z=0$, the only type of Einstein static solution possible in this context is the one in which the theory is effectively coincident with General Relativity (with the difference that in this solution $k=0$). The presence of $\bar{\mathcal{D}}$ and its features are consistent with the results of \cite{Boehmer:2013oxa}. In particular, we see that the static solutions can, differently from the case of General Relativity, be characterized by an non closed geometry.

The point $\bar{\mathcal{D}}$ is unstable for any values of the parameter $n$, but it is worth looking at the details of its stability in a more careful way. As we have mentioned in the previous section, the stability in terms of the barred variables presents some differences with respect to the one in the non barred variables. One of the reasons behind this difference is that when $\bar{X}<0$, effectively $\h\propto-H$ so that the dimensionless time $\mathcal{M}$ goes {\em backwards} with respect to $\mathcal{M}$. Remarkably, however, (and as expected on the physical point of view) such change does not affect the general structure of the phase space in the sense that the stability of the fixed points, and particularly the attractors, is generally not affected by the properties of the new time parametrisation. 

The exception is  for point $\bar{\mathcal{D}}$ in the interval $0<n<1$. Since in this interval this point is a repeller the correction for the time parameter would make ti an attractor for $\bar{X}<0$ and a repeller for $\bar{X}>0$. The consequence fo this result is that orbits will cross the $\bar{X}<0$ {\em through} the point $\bar{\mathcal{D}}$. Therefore in this specific case the presence of $\bar{\mathcal{D}}$ implies the presence of bounces, at least in the neighbourhood of this fixed point.
 
\begin{table}[tbp] \centering
\caption{The fixed points, their associated solutions  and stability for $f=\chi\R^n$ analyzed with the barred variables. Here $A$ stays for attractor, R for repeller, $S$ for saddle,F$_A$ for attractive focus and F$_R$ for repulsive focus. 
}
\begin{tabular}{cclllccc}
& & \\
\hline   Point &$(\bar{X}, \bar{Y}, \bar{Z}, \bar{\Omega},\bar{K})$ &  Scale Factor  & Energy Density &Stability\\ \hline
& & \\
$\bar{\mathcal{A}}$ &$\left\{1,2,\frac{2}{n-2},0,0\right\}$&  $a=a_0 \exp\left[\lambda(t-t_0)\right]$ & $\mu=0$ &F$_A$  \\ 
& & \\
\multirow{3}{*}{$\bar{\mathcal{B}}$} & \multirow{3}{*}{$\left\{\frac{1}{2-n},1- \frac{1}{(n-2)},0,0, -\frac{1}{(n-2)^2}\right\}$}&  \multirow{3}{*}{$a=a_0 \left(t-t_0\right)$} &\multirow{2}{*}{$\mu=\mu_0(t-t_0){}^{-3 (w+1)}$}& F$_A$~~$1<n<2$ \\ 
&&&&~~~~~ $0\leq w\leq 1$ \\
&&&& S otherwise\\
& &\\
\multirow{2}{*}{$\bar{\mathcal{C}}$} & $\left\{\frac{2}{-3 n (w+1)+3 w+5}, \frac{3 n (w+1)-4}{3 n (w+1)-3 w-5},\right.$&  \multirow{2}{*}{ $a=a_0\left(t-t_0\right)^{\frac{2}{3(1+w)}}$}&  \multirow{2}{*}{$\mu=\mu_0 \left(t-t_0\right)^{-2}$}& \multirow{2}{*}{S} \\ 
& $\left.0,\frac{4}{(-3 n (w+1)+3 w+5)^2}, -1\right\}$\\
& & \\
\multirow{2}{*}{$\bar{\mathcal{D}}$} & \multirow{2}{*}{$\left\{0,\frac{n}{n-1},0,0, 0\right\}$}&  \multirow{2}{*}{$a=a_0$} & \multirow{2}{*}{$\mu=0$} & R~~$0<n<1$  \\ 
&&&& S otherwise\\
& &\\
 \hline
\end{tabular}\label{T-Rn-PF_Bar}
\end{table}
\subsection{The hybrid HS model in the barred variables}
Let us analyze then the case of the hybrid HS model. The system in this case reads
\begin{align}
\begin{split}
& \bar{X}_{,\M}=X\left[K-Y+1-\frac{1}{2} (1+3 w) X\right] -\frac{1}{2} \{K (1+3 w)+3 (1+w) Z-\frac{Z^2}{\alpha Y^2}[(1-3 w) (K+1)+(1+3 w)Y]  \},\\
& \bar{Y}_{,\M}=2 Y\left[K+X-Y-\frac{ Z(X -1)}{2 Z-2 \alpha  Y}+1\right],\\ 
& \bar{Z}_{,\M}=2Z\left[(K+X-Y+1)- \frac{Z(X-1)}{2 Z-2 \alpha  Y}\right],\\
&\bar{K}_{,\M}=2\bar{K}(1+\bar{K}- \bar{Y}).
\end{split}
\end{align}
As in the previous section, the new system admits a set of fixed points that correspond to the ones of the formulation in terms of the variables \rf{DynVar} (see Table \ref{TableHSBar}) and,as before, in spite of he differences in the behaviour of the time variable, no differences in the stability of the fixed points arises.

In this case the Einstein static solution is represented by point $\bar{\mathcal{E}}$ which is characterized by a flat geometry and it is always a saddle. This suggests that the phenomenon observed in the previous example, i.e. the fact that bouncing orbits will pass through the Einstein point, is not present here, and the study of bouncing phenomenology requires numerical tools.

Finally, there is no global attractor which characterized by de Sitter solution. This confirms that the analysis of the behaviour of the action made in the beginning of  section \ref{HS} cannot be valid in general. 

\begin{table}[tbp] \centering
\caption{The fixed points and the solutions of the case with $f=\alpha \frac{\R}{1+\beta\R}$ in the barred varibales. The value of the parameter $\lambda$ is determined by \rf{LambdaValuesHS} and the coordinate of the fixed points.  Here $A$ stays for attractor, R for repeller, $S$ for saddle, F$_A$ for attractive focus and F$_R$ for repulsive focus.}
\begin{tabular}{ccccl}
& & \\
\hline   Point &$(\bar{X}, \bar{Y}, \bar{Z}, \bar{\Omega},\bar{K})$ &  Scale Factor  & Energy Density & Stability \\ \hline
& & \\
 \multirow{2}{*}{$\bar{\mathcal{A}}$}&  \multirow{2}{*}{$\left[1,2,2 \alpha -2 \sqrt{\alpha  (\alpha +1)},0, 0\right]$}&  \multirow{2}{*}{$a=a_0 \exp\left[\lambda(t-t_0)\right]$} &  \multirow{2}{*}{$\mu=0$} & F$_A$~~$0<\alpha< 49/32$ \\
&&&& A otherwise\\
& & \\
$\bar{\mathcal{B}}$& $\left[1,2,2 \alpha +2 \sqrt{\alpha  (\alpha +1)},0, 0\right]$& $a=a_0 \exp\left[\lambda(t-t_0)\right]$ & $\mu=0$ &  F$_A$  \\
& & \\
 $\bar{\mathcal{C}}$ &$\left[\frac{1}{3},\frac{8}{9}, 0,0, -\frac{1}{9}\right]$&$a=a_0 \left(t-t_0\right)$& $\mu=\mu_0(t-t_0)^{-3(1+w)}$ & S \\ 
 & & \\
  $\bar{\mathcal{D}}$& $\left[\frac{1}{3 w+4},\frac{3 w+7}{6 w+8},0,\frac{1}{(3 w+4)^2}, 0\right]$& $a=a_0 \left(t-t_0\right)$ & $\mu=\mu_0(t-t_0)^{-3(1+w)}$ & S \\
& & \\
 $\bar{\mathcal{E}}$& $\left[0,\frac{1}{2},0,0,0\right]$& $a=a_0 $&$\mu=0$&S\\
   & & \\ \hline
\end{tabular}\label{TableHSBar}
\end{table}

 \section{Conclusions}\label{conclusion}
In this paper we have applied a dynamical system analysis to investigate the cosmological features of hybrid metric-Palatini theories of gravitation. Making full use of the relation between the metrics $\hat{g}_{\mu\nu}$ and $g_{\mu\nu}$, it is possible to write the cosmological equation in a particularly compact way. Using these equations it is relatively straightforward to set up a dynamical systems approach for a wide range of hybrid metric-Palatini  models. In the present paper, we have focused our work on two types of functions $f$: the first is a generic power of the Palatini curvature $\R$; the second has a form that is inspired by the Hu-Sawicki model for metric $f(R)$ gravity. 

In the first case, the phase space presents  fixed points that are associated to solutions already found, for example in \cite{Capozziello:2012ny}, with a different approach. The dynamical system analysis, however, clarifies their stability and gives information of the global behaviour of the cosmology. For example, it appears clear that the cosmology always presents an attractor ($\mathcal{A}$) associated to an exponential solution.  In terms of stability, it is worth remembering that $\mathcal{A}$ is not  a global attractor for the phase space, because of the structure of the dynamical system. 

In the second case, the new formulation of the cosmological equations allows the deduction of some interesting properties of the model and in particular provide bounds for the values of the parameters. In terms of the phase space, we have now two attractors related to de Sitter solutions. However those fixed points are not global attractors for the cosmology. In this respect, the  dynamical system analysis reveals the limitations that conclusions drawn of the behaviour of the action for small and large values of the Ricci scalar.  

Both the examples we considered present fixed points that correspond to the GR limits of these models i.e. are characterised by $Z=0$ ($f=0$). Such points are never stable, so in principle there is no value of the parameters and set of initial conditions which might lead to cosmic histories in which the cosmology becomes indistinguishable form GR. This is in fact a desired behaviour as we aim at obtaining cosmologies which are close to GR at early times and at late time depart from Einstein cosmology.

The analysis in the alternative ''barred variables'' allows the study of the stability of the Einstein static universe in the context of these theories. It turns out that both the models we have analysed contain Einstein static fixed points, which are unstable. Their stability however is complicated by the choice of the time variable $\mathcal{M}$: in the case $\bar{X}<0$ this parameter has an opposite sign with respect to $\mathcal{N}$ and the stability analysis must be corrected accordingly.  In the case of $f=\chi \R^n$ this leads to an interesting phenomenon in terms of bounce phenomenology:  for a specific interval of $n$ orbits close to the Einstein static fixed point cross the $\bar{X}=0$ line only at this fixed points. This result is important in terms for the detailed study of bounce dynamics  and will be the focus of future studies.

\section*{Acknowledgements}
SC  was supported by  the Funda\c{c}\~{a}o para a Ci\^{e}ncia e Tecnologia through project IF/00250/2013. FSNL acknowledges financial support of the Funda\c{c}\~{a}o para a Ci\^{e}ncia e Tecnologia through an Investigador FCT Research contract, with reference IF/00859/2012, and the grants EXPL/FIS-AST/1608/2013 and PEst-OE/FIS/UI2751/2014.

\end{document}